\documentclass[conference]{IEEEtran}
\IEEEoverridecommandlockouts
\usepackage[noadjust]{cite}
\usepackage{amsmath,amssymb,amsfonts}
\usepackage{algorithmic}
\usepackage{graphicx}
\usepackage{textcomp}
\usepackage{xcolor}
\usepackage{tabularx, makecell, multirow, balance, booktabs}
\usepackage[numbers]{natbib} 
\usepackage{tikz}
\usepackage[T1]{fontenc}
\usepackage{enumitem}
\usepackage{url}

\makeatletter 
\newcommand{\linebreakand}{%
  \end{@IEEEauthorhalign}
  \hfill\mbox{}\par
  \mbox{}\hfill\begin{@IEEEauthorhalign}
}
\makeatother 

\newlength{\RoundedBoxWidth}
\newsavebox{\GrayRoundedBox}
\newenvironment{GrayBox}[1][\dimexpr\linewidth-4.5ex]%
   {\setlength{\RoundedBoxWidth}{\dimexpr#1}
    \begin{lrbox}{\GrayRoundedBox}
       \begin{minipage}{\RoundedBoxWidth}}%
   {   \end{minipage}
    \end{lrbox}
    \begin{center}
    \begin{tikzpicture}%
       \draw node[draw=black,fill=black!10,rounded corners,%
             inner sep=2ex,text width=\RoundedBoxWidth]%
             {\usebox{\GrayRoundedBox}};
    \end{tikzpicture}
    \end{center}}

\def\BibTeX{{\rm B\kern-.05em{\sc i\kern-.025em b}\kern-.08em
    T\kern-.1667em\lower.7ex\hbox{E}\kern-.125emX}}
\begin{document}

\title{The Good, the Bad, and the (Un)Usable: \\A Rapid Literature Review on Privacy as Code}
\author{
\IEEEauthorblockN{Nicol\'{a}s E. D\'{i}az Ferreyra\IEEEauthorrefmark{1}, Sirine Khelifi\IEEEauthorrefmark{1}, Nalin Arachchilage\IEEEauthorrefmark{2} and Riccardo Scandariato\IEEEauthorrefmark{1}}
\IEEEauthorblockA{\IEEEauthorrefmark{1}\textit{Institute of Software Security}, \textit{Hamburg University of Technology}, Hamburg, Germany}
\IEEEauthorblockA{\IEEEauthorrefmark{2}\textit{School of Computing Technologies}, \textit{RMIT University}, Melbourne, Australia \\
\{nicolas.diaz-ferreyra, sirine.khelifi, riccardo.scandariato\}@tuhh.de}\{nalin.arachchilage\}@rmit.edu.au
}

\maketitle

\begin{abstract}



Privacy and security are central to the design of information systems endowed with sound data protection and cyber resilience capabilities. Still, developers often struggle to incorporate these properties into software projects as they either lack proper cybersecurity training or do not consider them a priority. Prior work has tried to support privacy and security engineering activities through threat modeling methods for scrutinizing flaws in system architectures. Moreover, several techniques for the automatic identification of vulnerabilities and the generation of secure code implementations have also been proposed in the current literature. Conversely, such \textit{as-code} approaches seem under-investigated in the privacy domain, with little work elaborating on (i) the automatic detection of privacy properties in source code or (ii) the generation of privacy-friendly code. In this work, we seek to characterize the current research landscape of \textit{Privacy as Code} (PaC) methods and tools by conducting a rapid literature review. Our results suggest that PaC research is in its infancy, especially regarding the performance evaluation and usability assessment of the existing approaches. Based on these findings, we outline and discuss prospective research directions concerning empirical studies with software practitioners, the curation of benchmark datasets, and the role of generative AI technologies.

\end{abstract}

\begin{IEEEkeywords}
privacy engineering, privacy as code, rapid literature review, usability, automated software engineering
\end{IEEEkeywords}

\section{Introduction} \label{sec:intro}

Privacy engineering has become an imperative part of modern software development due to the emergence of strict legal frameworks such as the General Data Protection Regulation (GDPR) in the European Union \cite{riva2020sok}. Overall, it involves multiple activities (e.g., privacy requirements elicitation and threat modeling) that, altogether, seek to (i) bring privacy to the forefront of the development pipeline and (ii) facilitate compliance with these legal provisions and data protection standards alike \cite{iwaya2023wild}. Nonetheless, it remains a challenging task for many developers who do not count on extensive privacy training from either a technical or a legal perspective \cite{aljeraisy2022laws, bednar2019engineering}. In turn, privacy requirements are often seen as an afterthought, resulting in system designs and architectures devoid of data protection measurements and prone to personal information breaches \cite{senarath2018developers}.

To revert this tendency, prior work has introduced a wide array of methods and tools that support developers on privacy engineering activities \cite{riva2020sok}. These include privacy-aware methodologies for risk management and threat modeling that facilitate the selection and operationalization of Privacy-Enhancing Technologies (PETs) such as differential privacy, federated learning, and homomorphic encryption \cite{boteju2023sok}. LINDDUN \cite{wuyts2020linddun} is one of the most popular of these methods, which consists of a catalog of privacy threats that can be easily identified at the \textit{architectural level} of an information system. That is, from an abstract representation of the key computing elements of the system under analysis and the information flows exchanged between them \cite{tuma2019flaws}. This approach, in which architectural models are leveraged to conduct privacy assessments, is followed by many other frameworks and standards such as ProPAN \cite{ferreyra2020pdp}, PRIPARE \cite{notario2015pripare}, and the ISO 27550~\cite{iso27550}. Still, these techniques cannot guarantee full coverage, as architectural representations are often produced manually by analysts and do not necessarily correlate with the actual implementation of the system \cite{tuma2019flaws}. Consequently, gaps in the identification of privacy flaws may emerge depending on the precision and granularity of these representations \cite{pattakou2018towards,galvez2018odyssey}.

\textit{\textbf{Motivation:}} Motivated by the limitations of architecture-centered methods, recent work started to explore techniques for the (semi) automatic identification of privacy properties in \textit{source code} with minimum human intervention \cite{2024_tang}. Such analysis techniques are already quite mature and well-investigated in the \textit{security} domain \cite{ami2024false}. For instance, static security analysis tools allow for the early identification of vulnerabilities by checking the syntax, logic, and structure of source code. Dynamic tools, on the other hand, do so by running the code and observing its behavior under specific user inputs and execution scenarios \cite{song2019sok}. Although the latter has been fairly investigated in the \textit{privacy} realm (though largely concentrated on Android) \cite{2022_zhao}, the former remains underexplored and calls for a systematic assessment of the state-of-the-art. Particularly, a deep dive into the challenges and limitations of static analysis tools for privacy would help outline future research directions. Furthermore, given (i) the increasing number of tools targeting secure code generation available in the current literature and (ii) recent advances in generative Artificial Intelligence (AI) technologies, a knowledge synthesis of existing approaches addressing privacy-friendly implementations would also be beneficial.

\textit{\textbf{Contribution and Research Questions:}} This work provides a quick yet actionable overview of \textit{Privacy as Code} (PaC) methods and tools documented in the current literature. That is, on approaches elaborating either on (i) the automatic detection of privacy properties in source code, or (ii) the automatic generation of privacy-friendly implementations. Particularly, we delve into the scope, technicalities, and limitations of state-of-the-art approaches through a rapid literature review. All in all, we aim to answer the following Research Questions (RQs):


\begin{itemize}
    \item \textbf{RQ1: What is the scope of the techniques implemented by PaC methods and tools?} Code generated by PaC tools should implement one (or more) well-known privacy design strategies, such as \textit{minimizing}, \textit{abstracting}, or \textit{hiding} users' personal information \cite{hoepman2018privacy}. Similarly, PaC analysis tools should be capable of identifying threats such as \textit{unawareness}, \textit{data disclosure}, or \textit{non-compliance} \cite{wuyts2020linddun}. Hence, to answer this RQ, we identify the privacy threats and design strategies addressed by the primary studies selected in this review.

    \item \textbf{RQ2: What are the core technical foundations of PaC methods and tools?} Abstract source code representations, such as call graphs or abstract syntax trees, are fundamental to static analysis tools, as they help identify common programming mistakes. To answer this RQ, we summarize the code representation methods employed in the PaC literature and the programming languages covered by them.
 \item \textbf{RQ3: What are the main challenges and limitations of PaC methods and tools? } Finally, we identify the issues and open challenges stemming from the current literature concerning the technical performance and usability of existing PaC approaches. Based on these findings, we elaborate and discuss a roadmap to guide future research endeavors around PaC methods and tools. 
\end{itemize}

\section{Background and Related Work}

\textbf{Privacy as Architecture:} As mentioned earlier, several methods have been proposed to assess privacy threats in software architectures \cite{pattakou2018towards,senarath2019will}. At their core, these methods often require a model of the system under scrutiny to identify privacy issues in the information flows across architectural components. In the case of LINDUNN \cite{wuyts2020linddun}, these models are expressed as Data Flow Diagrams (DFDs) depicting processes, data stores, external entities, and the data exchanged between them. Similar to STRIDE \cite{van2022descriptive} (a threat modeling methodology for security), LINDDUN defines a schema of prospective privacy threats types that may affect specific DFD elements (i.e., \textbf{L}inkability, \textbf{I}dentifiability, \textbf{N}on-repudiation, \textbf{D}etectability, information \textbf{D}isclosure, \textbf{U}nawareness, and \textbf{N}on-Compliance). Such a schema is used in combination with threat trees and misuse cases to elicit privacy threat scenarios and suitable PETs to mitigate them. Some approaches like PRIPARE \cite{notario2015pripare} are agnostic with regard to the modeling language and can be applied in a top-down fashion (i.e., to model privacy requirements). Others like ProPAN \cite{ferreyra2020pdp} and PriS \cite{kalloniatis2008addressing} operate on architectural requirements that must be specified either as goal models or structured descriptions of the system's context. In any case, these methods demand significant manual effort and expertise that average analysts may lack \cite{bednar2019engineering,galvez2018odyssey}.


\textbf{Privacy as Code:} The term PaC was coined by the Ethyca group in the 2022 Open Security Summit as they presented FIDES, an open-source toolkit for checking privacy compliance in source code \cite{lapiana2022privacyascode}. It embraces the same spirit as ``security as code'' does in the context of DevSecOps, namely, the continuous integration of security principles (e.g., controls, policies, and best practices) directly and seamlessly into the software development pipeline \cite{vakhula2023research,das2023security}. Although PaC seems a novel concept, prior investigations have actively contributed to it for quite some time. For instance, \citet{ferrara2018tailoring} provided some initial ideas on how static program analysis techniques could support the detection of sensitive data processing in software systems. Particularly, they propose extending these techniques with categories of sensitive data, sources, and leakage points. Alongside, \citet{2020_hjerppe} introduced an annotation schema for Java classes and functions to facilitate the automatic detection of privacy violations in code, whereas \citet{2023b_tang} do so with the help of regular expressions and a standardized representation of personal data flows. All these methods have expanded the body of knowledge in PaC, mainly from a technical perspective. However, their challenges, limitations, and usability gaps have not been systematically investigated nor discussed in the current literature to the best of our knowledge.

\section{Methodology} \label{sec:method}

\begin{figure}[!t]
\centering
\includegraphics[width=0.65\linewidth]{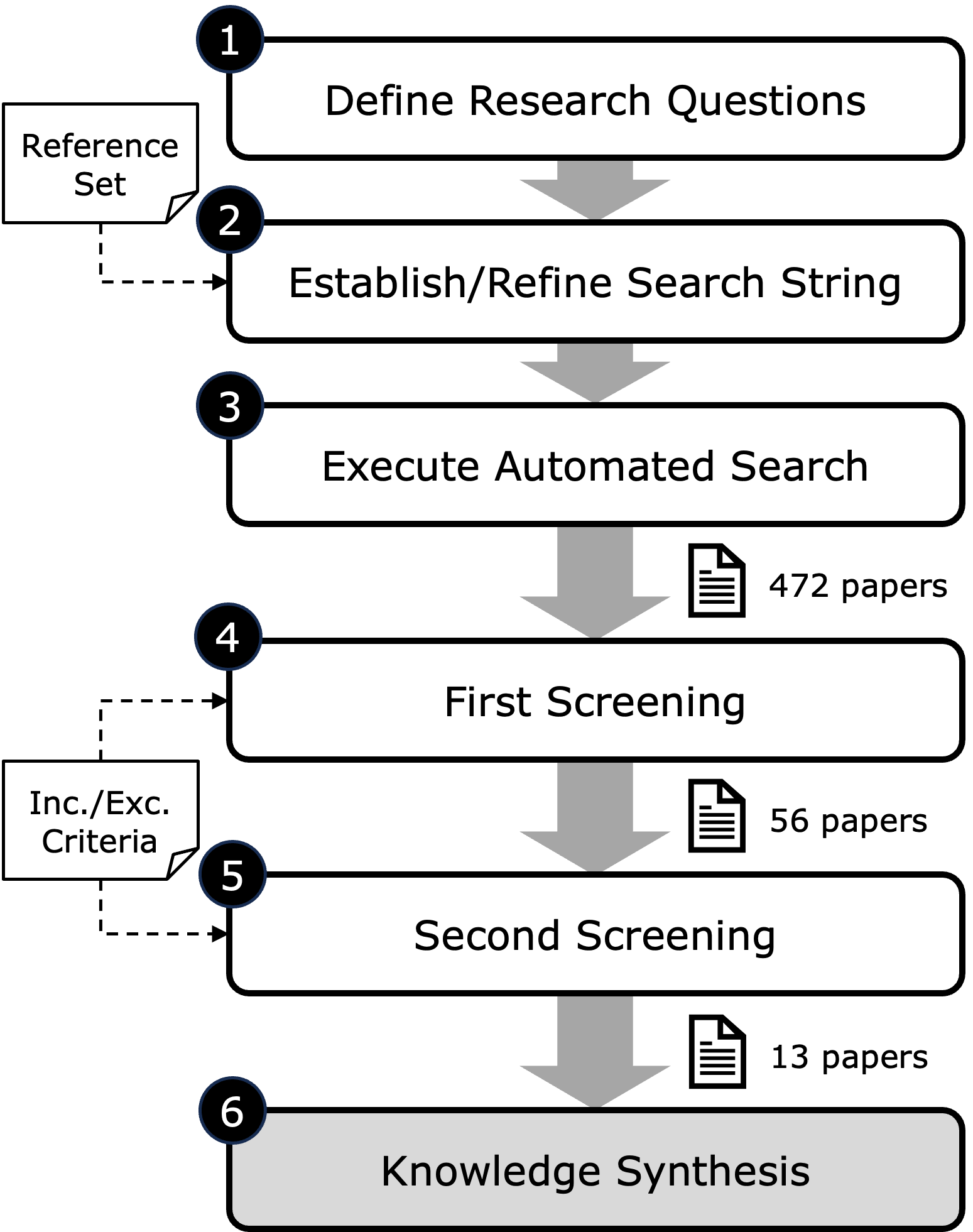}
    \caption{Applied Methodology: Rapid Literature Review.}
    \label{fig:method}
    \vspace{-2ex}
\end{figure}

To answer the RQs proposed in Section~\ref{sec:intro}, we conducted a rapid literature review on PaC methods and tools (Fig.~\ref{fig:method}). Rapid Reviews (RRs) are popular in the software engineering domain as they offer a quick overview of available evidence while omitting or simplifying some of the steps prescribed by systematic review frameworks (e.g., by using fewer databases and quality checks)~\cite{cartaxo2020rapid}. Thereby, they allow a timely yet actionable summarization of the state-of-the-art. In the following subsections, we describe the steps and materials used in this~study.

\subsection{Literature Search (Steps 1 to 3)}

\textit{Step 1} consists of defining the RQs to be answered by the RR, which, in this case, correspond to those presented in Section~\ref{sec:intro}. Next, we selected the engine and the string of keywords for the automated search of sources (\textit{Step 2}). We chose Scopus\footnote{\url{https://www.scopus.com/}} as our only search engine as it collects up-to-date information from some of the most relevant software engineering libraries (e.g., ACM and IEEE Xplore). For defining and streamlining the search string, we opportunistically built a reference set of papers (i.e., \cite{2020_hjerppe} and \cite{2023_kunz}) addressing PaC. Such a reference set was used to test different combinations of search terms while ensuring relevant results (both papers should be included in the search output). After some iterations, we agreed on the query depicted in Fig.~\ref{fig:query} and proceeded to process the literature items obtained from the automated search (\textit{Step 3}), which account for 472 references in total.

\subsection{Data Extraction (Steps 4 to 6)}

We assessed the relevance of each reference gathered in \textit{Step~3} based on the following inclusion/exclusion criteria:
\begin{itemize}
    \item \textit{Inclusion criteria}: Studies should describe privacy methods or tools that work at the code level. That is, either by (i) analyzing privacy threats in source code or (ii) generating privacy-friendly implementations. They should also achieve a certain level of automation in both cases and be published in peer-reviewed outlets (i.e., journals, conferences, or workshops). To further reduce the scope of our analysis, we focused exclusively on studies dated 2016 or later, namely, after the European Union formally adopted the GDPR.
    \item \textit{Exclusion criteria}: Studies addressing privacy solely on an architectural level were excluded. Same with papers not written in the English language and published before 2016. Grey literature, such as blog posts, white papers, and opinion pieces, was not taken into consideration. Sources from low-quality outlets (e.g., unknown venues) were also removed to reach a high standard of scientific~evidence.
\end{itemize}

These criteria were first applied to the title and abstract of each of the 472 sources (\textit{Step 4}), resulting in 56 papers that were then scrutinized in full (\textit{Step 5}). At the end of this process, we retained 13 sources (see Table~\ref{tab:selected}) that underwent a knowledge synthesis process (\textit{Step 6}). For this, we followed a lightweight open coding approach in which one author extracted emerging themes and patterns from the data under the lens of the proposed RQs. Another author reviewed these themes afterward and, if discrepancies arose, these were resolved through a negotiated agreement.

\begin{figure}[!h]
\noindent\fbox{%
    \parbox{0.98\linewidth}{\centering\small
(``\texttt{code}''<OR>``\texttt{tool}''<OR>``\texttt{static analysis}'') <AND>\\
(``\texttt{privacy analysis}''<OR>``\texttt{privacy protection}''\\<OR>``\texttt{data privacy}'')\\
<AND> (``\texttt{compliance}''<OR>``\texttt{GDPR}'')
    }%
}
\caption{Final search query.}\vspace{-2ex}
\label{fig:query}
\end{figure}

\section{Results and Discussion}

The RR study was conducted in April 2024. In the next subsections, we report the findings of the knowledge synthesis process (summarized in Table~\ref{tab:classification}). All study materials, including the code book and annotated spreadsheets, are publicly available in the paper's \textbf{Replication Package}\footnote{\url{https://doi.org/10.5281/zenodo.14671296}}.

\subsection{Scope and techniques (RQ1)}


\subsubsection{Code analysis}  We observed that most PaC approaches in the literature focus on the analysis of privacy properties in source code (11 sources). Furthermore, they often apply some sort of static taint analysis technique to detect streams of confidential information flowing (either explicitly or implicitly) out of the system's trust boundaries (e.g., to the Internet) \cite{2018b_ferrara}. For this, these techniques must first identify which types of sensitive data and leakage points exist in the system and how such data can be accessed (and leaked) across different program statements (e.g., method calls). Then, by reconstructing the control flows of the program, they can spot \textit{tainted} execution paths in which sensitive data reach a leakage point without proper sanitization (e.g., non-encrypted). 

Normally, it is the user of PaC tools (e.g., a privacy analyst) the one responsible for identifying (i) the personal data items in the code and (ii) the program statements processing them (often referred to as \textit{sources} and \textit{sinks}, respectively). This task is typically supported through annotation schemas and guidelines provided with the tools for the sake of consistency. For instance, \citet{2020_hjerppe} defined a set of rules and tags for labeling Java classes processing personal data, whereas \citet{2018b_ferrara} prescribed a categorization of sensitive data and leakage points to annotate API methods and parameters. Another approach by \citet{2023_ferreira} requires a user-defined list of personal data in the form of database tables and column names, while in some other cases, this is done in a more automated fashion through advanced NLP techniques. Such is the case of \citet{2022_grunewald}, who applied Named Entity Recognition (NER) to create inventories of personal data stored across distributed web services, or \citet{2022_zhao}, who used a Bi-directional Long Short Term-Memory (LSTM) mechanism together with similarity metrics to map privacy policies to code variables.




When analyzed under the lens of LINDDUN, we can see that the scope of PaC analysis tools typically covers \textit{data disclosure} and \textit{non-compliance} threats, which is achieved (to a large extent) through the identification of unforeseen and unauthorized flows of personal data in the system's code. Although a thorough GDPR compliance check would also require a definition of the corresponding privacy policies and relevant legal provisions (e.g., like in \cite{2023_ferreira}), most approaches contribute to maintaining a record of the data processing activities in the software under scrutiny (Article 30). Other methods, like the one proposed by \citet{2021_grunewald}, address \textit{unawareness} threats more explicitly through transparency vocabularies and domain-specific languages used to delimit the purpose and retention periods of data processing activities. Further information about the presence of pseudo-identifiers in database operations (i.e., read and write queries) can also support the identification of additional privacy threats such as \textit{linkability}, as shown by \citet{2023_kunz}.




 \begin{table}[!t]
\def\arraystretch{0.75}
\caption{Primary studies identified during the review.}
\label{tab:selected}
\scriptsize \centering
\begin{tabularx}{\linewidth}{lcccc}
\toprule
\textbf{Source} & \textbf{Year} & \textbf{Type} & \textbf{Approach} & \textbf{OS Tool} \\
\midrule
\citet{2018b_ferrara} & 2018 & conference & analyze & NO\\       
\midrule  
\citet{2020_hjerppe} & 2020 & conference & analyze & YES\\ 
\midrule
\citet{2021_grunewald} & 2021 & workshop & analyze & YES\\
\midrule  
\citet{2022_grunewald} & 2022 & conference & analyze & YES \\    
\midrule  
\citet{2022_pallas} & 2022 & conference & generate & YES\\    
\midrule                
\citet{2022_zhao} & 2022 & conference & analyze & YES\\       
\midrule 
\citet{2023_ferreira} & 2023 & conference & analyze & YES\\   
\midrule  
\citet{2023b_tang} & 2023 & conference & analyze & NO\\         
\midrule 
\citet{2023_hjerppe} & 2023 & journal & analyze & YES\\ 
\midrule
\citet{2023_kunz} & 2023 & journal & analyze & NO\\ 
\midrule  
\citet{2023_goldsteen} & 2023 & journal & generate & YES\\   
\midrule  
\citet{2024_tang}& 2024 & workshop & analyze & NO\\ 
\midrule  
\citet{2024_morales} & 2024 & conference & analyze & NO\\       
\bottomrule
\end{tabularx}
\vspace{-2ex}
\end{table}
 
\subsubsection{Code generation} Only two PaC generation techniques were identified among the selected primary studies. One, proposed by \citet{2022_pallas}, consists of a collection of schema directives that enforce data minimization strategies in GraphQL web APIs. Such directives define post-processing steps (e.g., hashing) to be applied on precomputed data fields before returning them through the API. This approach is available as a plug-in for Apollo, a toolkit for building GraphQL solutions, and can be extended with custom access-control policies and information reduction methods. The second technique, by \citet{2023_goldsteen}, also addresses data minimization but in the context of Machine Learning (ML) projects and data-centric development pipelines. Like in the previous case, it implements data minimization strategies that can help reduce the amount of personal data needed to perform ML predictions. However, it also defines methods for anonymizing the training data of ML models while minimizing the impact on the model's performance. For this, a surrogate model is trained and used to identify (and generalize) groups of samples that behave similarly. All in all, this approach and the one of \citet{2022_pallas}, narrow their scope to techniques for \textit{minimizing}, \textit{hiding}, or \textit{generalizing} (abstracting) personal information in source code implementations.










\subsection{Technical foundations (RQ2)}

\begin{table}[!t]
\def\arraystretch{1.0}
\caption{Classification of the primary studies.}
\label{tab:classification}
\scriptsize \centering
\begin{tabularx}{\linewidth}{llc}
\toprule 
\textbf{Aspect} & \textbf{Approach} & \textbf{Sources} \\
\midrule
\multirow{2}{*}{Technique}
& Static taint analysis          & \cite{2018b_ferrara,2020_hjerppe,2023_hjerppe, 2022_zhao,2023_ferreira,2023b_tang, 2023_kunz,2024_morales,2024_tang}    \\
& Other        & \cite{2021_grunewald,2022_grunewald,2023_goldsteen,2022_pallas}    \\       
\midrule
\multirow{6}{*}{\makecell[l]{Abstraction}}
& Abstract Syntax Trees         & \cite{2020_hjerppe,2023_hjerppe, 2022_zhao,2023b_tang,2024_morales,2024_tang}   \\
& Code Property Graphs        &  \cite{2023_kunz} \\
& Call Graphs & \cite{2022_zhao,2024_tang} \\
& Bytecode        & \cite{2024_morales}   \\ 
& Other       & \cite{2018b_ferrara,2023_ferreira,2024_morales}   \\ 
& None & \cite{2021_grunewald,2022_pallas,2023_goldsteen}\\
\midrule
\multirow{6}{*}{\makecell[l]{
Languages}}
& Java & \cite{2018b_ferrara,2020_hjerppe,2023b_tang,2024_tang, 2023_hjerppe}    \\
& Javascript  & \cite{2023_ferreira,2023b_tang,2024_tang}  \\
& Python  & \cite{2022_zhao,2023_kunz,2023_goldsteen}    \\ 
& Go     & \cite{2023_kunz}    \\ 
& TypeScript     & \cite{2023b_tang}   \\ 
& Other (query, spec.)     & \cite{2022_grunewald,2021_grunewald,2022_pallas,2024_morales}  \\ 
\midrule
\multirow{4}{*}{\makecell[l]{Privacy\\ threats}}
& Data disclosure        & \cite{2021_grunewald,2022_grunewald,2018b_ferrara,2020_hjerppe, 2023_hjerppe, 2022_zhao,2023_ferreira,2023b_tang,2023_kunz,2024_morales,2024_tang}  \\
& Non compliance         & \cite{2021_grunewald,2022_grunewald,2018b_ferrara, 2020_hjerppe, 2023_hjerppe, 2022_zhao,2023_ferreira,2023b_tang,2023_kunz,2024_morales,2024_tang}   \\ 
& Unawareness        & \cite{2021_grunewald,2023_kunz,2023_ferreira}    \\
& Other       & \cite{2023_kunz}   \\          
\midrule                
\multirow{3}{*}{\makecell[l]{Privacy\\ design\\ strategies}}
& Minimize        & \cite{2022_pallas,2023_goldsteen}   \\
& Abstract         & \cite{2022_pallas,2023_goldsteen}  \\ 
& Hide        & \cite{2022_pallas}  \\ 
\bottomrule
\end{tabularx}
\vspace{-3ex}
\end{table}

As mentioned earlier, source code abstractions play a major role in static analysis tools targeting the identification of security issues, and it is not the exception for PaC ones. As shown in Table~\ref{tab:classification}, most PaC analysis tools use abstract representations of the source code under analysis to spot personal data leakage within program statements. Among them, Abstract Syntax Trees (ASTs) are the most popular ones with 5 out of 11 approaches adopting it as part of their technical foundations. At their core, ASTs represent and organize a code's syntax (e.g., statements, expressions, and control structures) into the nodes of a tree-like structure. ASTs are general-purpose data structures that do not focus on privacy per se. Hence, PaC methods either (i) enrich them with privacy properties or (ii) use them in tandem with other code abstractions to support the analysis of personal/sensitive information flows. For instance, \citet{2023b_tang} add personal data tags to ASTs using regular expressions and a predefined list of identifiers, whereas \citet{2022_zhao} use Call Graphs (CGs) to obtain the corresponding privacy delivery path (i.e., how information is exchanged across subroutines). 

Other abstractions such a Code Property Graphs (CPGs) \cite{2023_kunz} and Java bytecode \cite{2024_morales} have also been employed in the PaC literature. The former consists of DFDs enhanced with taint labels extracted from code annotations or comments (e.g., \texttt{@Identifier}) indicating the presence of (pseudo-) identifiers and database operations. In the case of the latter, it corresponds to an intermediate low-level representation of Android application code which is packaged inside APK files along with other resources. Such a representation is used to statically examine the information flows in Android source code with the help of third-party tools like FlowDroid \cite{arzt2014flowdroid}. From Table~\ref{tab:classification}, we observe that most PaC tools focus on code written in Java/Javascript followed by applications implemented in Python. Some approaches also provide evidence about their applicability to other programming languages like Go (e.g., \cite{2023_kunz}) or TypeScript (e.g., \cite{2023b_tang}) but, in most cases, tools tend to concentrate on a single language. Other techniques addressing web services seek for privacy-related statements inside GraphQL queries \cite{2022_pallas}, Infrastructure as Code files \cite{2022_grunewald}, and OpenAPI specifications \cite{2021_grunewald}.




\subsection{Challenges and limitations (RQ3)} \label{sec:challenges}

Like in the security domain, PaC tools struggle to achieve high levels of performance and scalability. Whereas some approaches are capable of checking systems consisting of several lines of code and dependencies (e.g., \cite{2023_kunz, 2022_zhao}), their scalability is mainly impaired by the relatively high number of false positive warnings they produce. This translates into additional efforts from developers and code reviewers to exclude non-relevant results, namely information flows that may seem privacy-relevant at a glance, but actually are not \cite{2023_ferreira, 2023_kunz}. Such a limitation is particularly challenging when parsing code written in JavaScript or Python as variables can change their types at runtime \cite{staicu2019empirical}. On the other hand, the coverage of PaC tools is not perfect and highly dependent on predefined lists of ``items of interest'' (e.g., personal data types) and the precision of the rules (e.g., regular expressions) or the tools (e.g., FlowDroid) used for their identification. In turn, relevant sinks, sources, and data flows may be overlooked, significantly affecting the reliability of these techniques \cite{2024_tang}.

\begin{figure}[t]
\centering
\includegraphics[width=\linewidth]{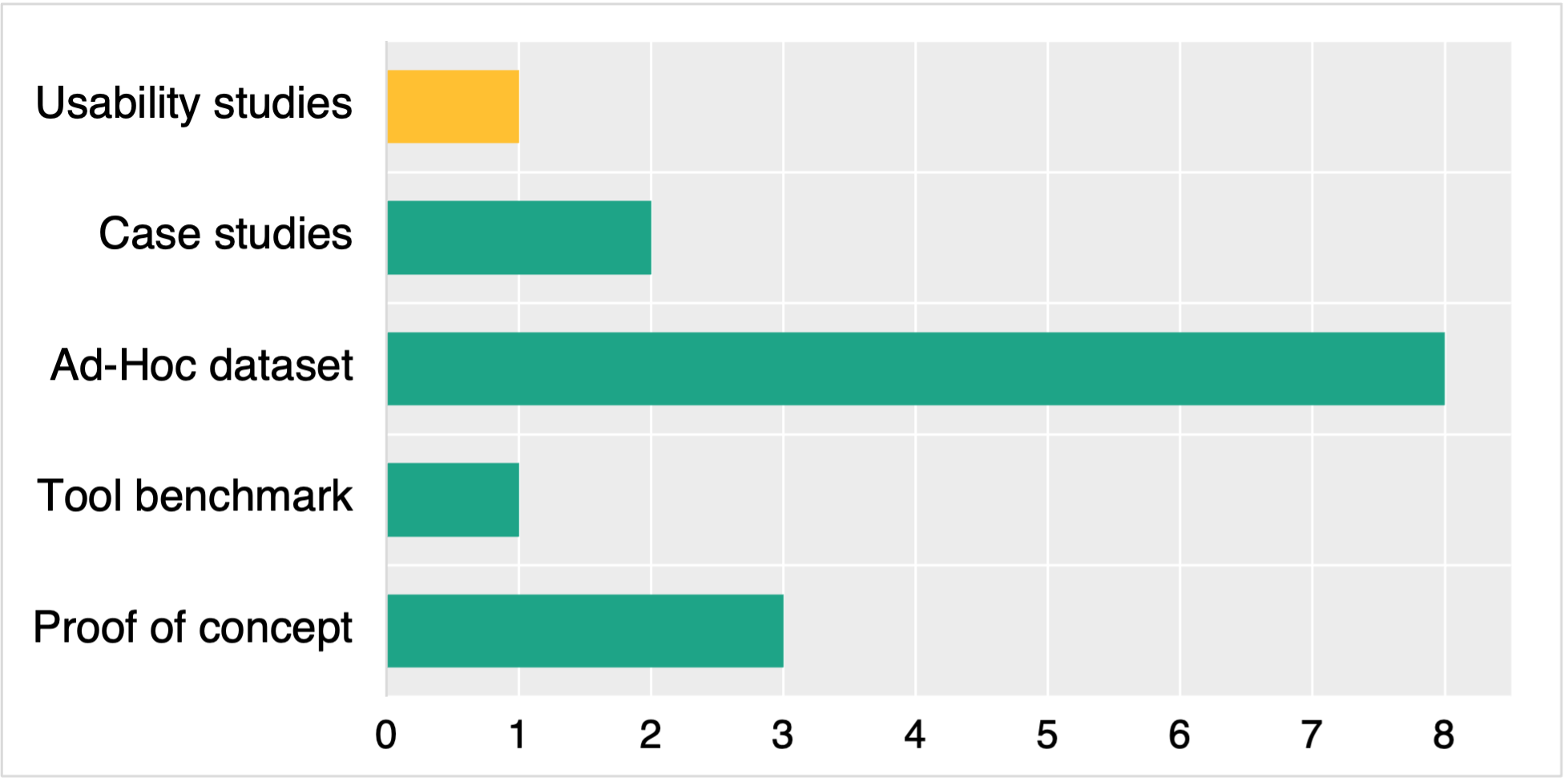}
    \caption{Evaluation strategies of PaC approaches.}
\label{fig:evaluation}
\vspace{-2ex}
\end{figure}

In terms of scope, there is a clear lack of PaC tools that (i) focus on the automatic generation of privacy-compliant code, and (ii) can run across multiple programming languages. Large Language Models (LLMs) like GPT-4 or LLaMa could help bridge these issues thanks to their advanced NLP capabilities. In fact, LLMs are being actively investigated to support the automatic generation of security-aware code from NL descriptions (e.g., \cite{black2024balancing}). Hence, many of these approaches could be, in principle, extrapolated to the privacy domain with minimum effort (e.g., by crafting privacy-aware code generation/analysis prompts). Still, privacy assessments are highly influenced by contextual factors (e.g., social norms or culture) that current LLM technologies struggle to navigate \cite{brown2022does}. This could, for instance, lead to incomplete or biased privacy judgments resulting in critical data flows being systematically ignored. Certainly, current PaC approaches may also make such omissions. However, the privacy assumptions behind LLMs are much more difficult to expose than those of the state-of-the-art methods, which are often encoded as rules (e.g., regular expressions) easy to scrutinize. Therefore, transparency and explainability will become major challenges for the next generation of LLM-based PaC solutions.








Conducting in-depth evaluations of PaC methods is still an open challenge, given the absence of well-established ground truth datasets \cite{2023b_tang}. As depicted in Fig.~\ref{fig:evaluation}, most methods employ ad-hoc datasets for their evaluation  \cite{2020_hjerppe,2022_zhao,2022_grunewald,2023b_tang,2023_goldsteen,2023_kunz,2024_morales,2024_tang} or elaborate on case studies \cite{2023_ferreira,2023_hjerppe}, while others are simply reduced to conceptual demonstrations \cite{2018b_ferrara,2021_grunewald,2022_pallas}. Hence, the performance of these tools is not measured properly outside their specific testing conditions, leaving many questions about their generalizability unanswered. Moreover, only one source compares their results against other (commercial) methods alike \cite{2022_grunewald}, whereas another one conducts a usability study of the proposed tool \cite{2023_ferreira}. Considering the significant amount of manual work these approaches require (e.g., code annotations), a thorough assessment of their operability and ease of use becomes critical. This should be fostered through an Open Source (OS) culture across PaC researchers in which they are encouraged to make their tools publicly available for further investigations (e.g., empirical studies with software practitioners).

\begin{figure}[t] \label{fig:outlook}
\begin{GrayBox}\small
\textbf{PROSPECTIVE RESEARCH PATHWAYS.}
\vspace{0.4ex}
\begin{enumerate}[leftmargin=3ex]
\item Curate ground-truth datasets to facilitate in-depth evaluations of PaC methods and tools while fostering an OS culture.\vspace{1ex}
\item Investigate the applicability of LLMs to PaC along with their potential transparency and explainability drawbacks.\vspace{1ex}
\item Assess the usability of PaC tools through thorough empirical studies with practitioners of diverse backgrounds. 
\end{enumerate}
\end{GrayBox}\vspace{-3ex}
\end{figure}

\section{Study Limitations}

A single database (Scopus) was used in this study, which may have limited to some extent the number of reviewed sources. Furthermore, the proposed inclusion/exclusion criteria and the search query could also have constrained the selection of relevant PaC approaches. Nevertheless, the application of a forward snowballing iteration and the use of a reference set helped us to mitigate this threat. Regarding the screening and knowledge synthesis process, these were initially conducted by a single researcher. As described in Section~\ref{sec:method}, another author closely reviewed the outputs of these steps to avoid confirmation and interpretation biases, respectively. 

\section{Conclusion and Future Work}
All in all, PaC is a paradigm that promises to make privacy-by-design principles more accessible and actionable to software practitioners. Still, current approaches cannot guarantee full compliance on their own and exhibit significant limitations in terms of scope, performance, and usability. Moreover, although PaC's promise of (full) automation seems appealing, one must bear in mind that privacy requires a thorough assessment of the context in which (personal) information flows across. In principle, this suggests that human judgment cannot be completely factored out from the development of privacy-friendly code despite the recent advances in Generative AI technologies. Instead, future PaC endeavors should actively support practitioners with methods and artifacts that help them navigate the complexities of contextual variables (e.g., trust assumptions, data sensitivity, and social norms). Therefore, large-scale evaluations of current and upcoming approaches are imperative in light of the sociotechnical factors influencing their adoption and technical soundness.












\bibliographystyle{IEEEtranN}

{\footnotesize
\bibliography{references.bib}}

\end{document}